\documentclass[10pt]{article}
\usepackage{amssymb}
\usepackage{amsmath}
\usepackage{graphicx}
\usepackage{fancyhdr}
\usepackage{cite}

\begin{document}

\newcommand{\bin}[2]{\left(\begin{array}{c}\!#1\!\\\!#2\!\end{array}\right)}

\huge

\begin{center}
On the Li-Rosmej analytical formula for energy level shifts in dense plasmas
\end{center}

\vspace{0.5cm}

\large

\begin{center}
Jean-Christophe Pain\footnote{jean-christophe.pain@cea.fr}
\end{center}

\normalsize

\begin{center}
CEA, DAM, DIF, F-91297 Arpajon, France
\end{center}

\vspace{0.5cm}


\begin{abstract}
Li and Rosmej derived analytical fits for the energy level shifts due to plasma screening on the basis of a free-electron potential published by Rosmej \emph{et al.} one year earlier. The derivation of the fits, which were shown by Iglesias to be inconsistent with the fundamental premise of the ion-sphere model, was motivated by the belief that no analytical expression exists for the expectation value $\langle r^{3/2}\rangle$, an assertion that was also contradicted by Iglesias. In this short note, I point out that a simple expression for the latter quantity can be obtained as a particular case of a formula published by Shertzer, and I provide a corresponding compact analytical expression for the level shifts in the framework of Rosmej's formalism.  
\end{abstract}

\section{Introduction}

In order to take into account plasma density effects on bound energy levels, several analytical formulas were obtained (see for instance the non-exhaustive list of references \cite{MASSACRIER90,LI12,POIRIER15,BELKHIRI15}) to approach ion-sphere potentials. Applying first-order perturbation theory to them together with hydrogenic-scaled mean ionization yields analytical formulas to estimate level shifts. In 2011, Rosmej \emph{et al.} proposed an asymptotic expansion to express the free-electron screening potential in finite-temperature plasmas in a closed analytical expression \cite{ROSMEJ11} (in atomic units):

\begin{equation}\label{ros}
V_{\mathrm{f}}(r)=4\pi n_e\left(R_{\mathrm{ws}}\right)\left\{\frac{R_{\mathrm{ws}}^2}{2}-\frac{r^2}{6}+\frac{4}{3\sqrt{\pi}}\left[\frac{\left(Z-N_{\mathrm{b}}\right)}{k_BT_e}\right]^{1/2}R_{\mathrm{ws}}^{3/2}-\frac{8}{15\sqrt{\pi}}\left[\frac{\left(Z-N_{\mathrm{b}}\right)}{k_BT_e}\right]^{1/2}r^{3/2}\right\},
\end{equation}

\noindent where $n_e\left(R_{\mathrm{ws}}\right)$ is the free-electron density at the Wigner-Seitz radius (radius of the ion-sphere), $N_{\mathrm{b}}$ the number of bound electrons, $k_B$ the Boltzmann constant and $T_e$ the electron temperature. We have

\begin{equation}
R_{\mathrm{ws}}=\left[\frac{3\left(Z-N_{\mathrm{b}}\right)}{4\pi n_e\left(R_{\mathrm{ws}}\right)}\right]^{1/3}.
\end{equation}

The energy shift of $n\ell$ subshell is then obtained by

\begin{equation}\label{delt}
\Delta\epsilon_{n\ell}=\langle n\ell|V_{\mathrm{f}}(r)|n\ell\rangle=\int_0^{\infty}V_{\mathrm{f}}(r)R_{n\ell}^2\left(r;Z_{\mathrm{eff}}\right)r^2dr,
\end{equation}

\noindent where $R_{n\ell}\left(r;Z_{\mathrm{eff}}\right)$ represents the radial part of the hydrogenic wavefunction of the subshell with effective nuclear charge $Z_{\mathrm{eff}}$. The formula involves quantities such as

\begin{equation}\label{squ}
\langle r^2\rangle=\frac{n^2}{2Z_{\mathrm{eff}}^2}\left[5n^2+1-3\ell(\ell+1)\right]
\end{equation}

\noindent and $\langle r^{3/2}\rangle$ (I use the simplified notation $\langle n\ell|f(r)|n\ell\rangle=\langle f(r)\rangle$). In 2012, Li and Rosmej, convinced that no formula exists for $\langle r^{3/2}\rangle$, derived an alternative analytical fit for $V_{\mathrm{f}}(r)$, depending only on $\langle r^2\rangle$ and on $\langle r\rangle$, which is known to be:

\begin{equation}
\langle r\rangle=\frac{1}{2Z_{\mathrm{eff}}}\left[3n^2-\ell(\ell+1)\right].
\end{equation}   

\noindent It was shown very recently by Iglesias that the fit published by Li and Rosmej was inconsistent with the ion-sphere model \cite{IGLESIAS19}, on the contrary to the ``original'' potential of Eq. (\ref{ros}). In addition, the remark of Li and Rosmej, which led to the work of Ref. \cite{LI12}, is not correct: $\langle r^{3/2}\rangle$ can definitely be expressed analytically. This was also pointed out in Ref. \cite{IGLESIAS19}, where the author indicates that such a quantity can be evaluated analytically, following the procedure given in Appendix E of Ref. \cite{SZMYTKOWSKI97} using the generating-function formalism (see for instance the textbook by Bransden and Joachain \cite{BRANSDEN83}) and yielding to a complicated expression (in the same paper, a table is provided with particular values displayed in the form of rational fractions). I would like to mention here that a simple expression for $\langle r^{3/2}\rangle$ exists. It is a particular case of a relation published by Shertzer in 1991 \cite{SHERTZER91}, who provided  an expression for $\langle n\ell|r^{\beta}|n\ell'\rangle$ for arbitrary $\beta$:

\begin{equation}\label{she}
\langle n\ell|r^{\beta}|n\ell'\rangle=A_{n,\ell,\ell'}\sum_{i=0}^{n-\ell-1}\frac{(-1)^i\Gamma(\ell+\ell'+3+i+\beta)}{i!(2\ell+1+i)!(n-\ell-i-1)!}\frac{\Gamma(\ell-\ell'+2+i+\beta)}{\Gamma(\ell+3-n+i+\beta)}
\end{equation}

\noindent and

\begin{equation}
A_{n,\ell,\ell'}=\frac{(-1)^{n-\ell'-1}}{2n}\left(\frac{n}{2Z_{\mathrm{eff}}}\right)^{\beta}\left[\frac{(n+\ell)!(n-\ell-1)!}{(n+\ell')!(n-\ell'-1)!}\right]^{1/2},
\end{equation}

\noindent applying therefore also for off-diagonal terms ($\ell\ne\ell'$). $\Gamma$ is the usual Gamma function, which evaluation was widely treated in the literature (see for instance the recent simple and efficient approximation given by Chen \cite{CHEN16}). In the present case, we have $\ell=\ell'$ and $\beta=3/2$, and we get

\begin{equation}\label{shej}
\langle r^{3/2}\rangle=\frac{(-1)^{n-\ell-1}}{2n}\left(\frac{n}{2Z_{\mathrm{eff}}}\right)^{3/2}\sum_{i=0}^{n-\ell-1}\frac{(-1)^i\Gamma(2\ell+9/2+i)}{i!(2\ell+1+i)!(n-\ell-i-1)!}\frac{\Gamma(7/2+i)}{\Gamma(\ell+9/2-n+i)}.
\end{equation}

Is is worth mentioning that a relativistic equivalent of Eq. (\ref{she}) for $\ell=\ell'$ was published by Salamin in 1995 \cite{SALAMIN95}. Inserting Eqs. (\ref{squ}) and (\ref{shej}) in Eq. (\ref{delt}) gives

\begin{eqnarray}
\Delta\epsilon_{n\ell}&=&\frac{\left(Z-N_{\mathrm{b}}\right)}{2R_{\mathrm{ws}}}\left\{3-\frac{n^2\left[5n^2+1-3\ell(\ell+1)\right]}{2R_{\mathrm{ws}}^2Z_{\mathrm{eff}}^2}+8\sqrt{\frac{\left(Z-N_{\mathrm{b}}\right)}{\pi R_{\mathrm{ws}}k_BT_e}}\left[1+\frac{(-1)^{n-\ell}\sqrt{n}}{5\left(2R_{\mathrm{ws}}Z_{\mathrm{eff}}\right)^{3/2}}\right.\right.\nonumber\\
& &\times\left.\left.\sum_{i=0}^{n-\ell-1}\frac{(-1)^i\Gamma(2\ell+9/2+i)}{i!(2\ell+1+i)!(n-\ell-i-1)!}\frac{\Gamma(7/2+i)}{\Gamma(\ell+9/2-n+i)}\right]\vphantom{\sqrt{\frac{\left(Z-N_{\mathrm{b}}\right)}{\pi R_{\mathrm{ws}}k_BT_e}}}\right\}.
\end{eqnarray}

The potential first published by Rosmej \emph{et al.} in 2011 \cite{ROSMEJ11} and which is consistent with the fundamental neutrality requirement of the ion-sphere model as shown by Iglesias \cite{IGLESIAS19}, can therefore be directly used to derive simple analytical formulas to estimate energy level shifts in dense plasmas. The alternative fit by Li and Rosmej \cite{LI12}, which is not consistent with the ion-sphere model, was motivated by the belief that no analytical expression exists for $\langle r^{3/2}\rangle$, a statement that was invalidated by Iglesias as well. In this short note, I pointed out that a compact analytical expression can be obtained for that quantity, as a particular case of a relation published by Shertzer, and the resulting formula for the energy level shift due do plasma screening effects was given, for a direct use in atomic-structure codes. 

\section*{References}



\begin{thebibliography}{99}

\bibitem{MASSACRIER90} G. Massacrier and J. Dubau, J. Phys. B: At. Mol. Opt. Phys. {\bf 23}, 24595 (1990).

\bibitem{LI12} X. Li and F. B. Rosmej, Euro. Phys. Lett. {\bf 99}, 33001 (2012).

\bibitem{POIRIER15} M. Poirier, High Energy Density Phys. {\bf 15}, 12 (2015).

\bibitem{BELKHIRI15} M. Belkhiri, C. J. Fontes and M. Poirier, Phys. Rev. A {\bf 92}, 032501 (2015).

\bibitem{ROSMEJ11} F. B. Rosmej, K. Bennadji and V. S. Lisitsa, Phys. Rev. A {\bf 84}, 032512 (2011).

\bibitem{IGLESIAS19} C. A. Iglesias, High Energy Density Phys. {\bf 30}, 41 (2019).

\bibitem{SZMYTKOWSKI97} R. Szymtkowski, J. Phys. B {\bf 30}, 825 (1997).

\bibitem{BRANSDEN83} B. H. Bransden and C. J. Joachain, {\it Physics of atoms and molecules} (Longman, Essex UK, 1993).

\bibitem{SHERTZER91} J. Shertzer, Phys. Rev. A {\bf 44}, 2832 (1991).

\bibitem{CHEN16} C.-P. Chen, J. Number Theory {\bf 164}, 417 (2016). 

\bibitem{SALAMIN95} Y. I. Salamin, Phys. Scr. {\bf 51}, 137 (1995).

\end{thebibliography}
\end{document}